\documentclass[]{aa}  

\usepackage{graphicx}
\usepackage{txfonts}
\usepackage{lipsum}
\usepackage{subcaption}         
\usepackage{lscape}             
\usepackage{placeins}           
                                
\begin{document}

\title{JWST detection of methane emission from a young brown dwarf}
\titlerunning{Methane emission from a young brown dwarf}


%
%
%

   \author{K. L. Luhman\inst{1,2}\corrauth{kll207@psu.edu},
C. Alves de Oliveira\inst{3},
J. K. Faherty\inst{4,5},
B. C. Smith\inst{6}}

\institute{Department of Astronomy and Astrophysics, The Pennsylvania State University, University Park, PA 16802, USA
\and
Center for Exoplanets and Habitable Worlds, The Pennsylvania State University,
University Park, PA 16802, USA
\and
European Space Agency, European Space Astronomy Centre, 
Camino Bajo del Castillo s/n, 28692 Villanueva de la Ca\~{n}ada, Madrid, Spain
\and
Museum of Natural History, New York, NY 10024, USA
\and
Department of Physics, The Graduate Center City University of New York, 
New York, NY 10016, USA
\and
Spectros Associates, 204 Wildflower, Ellicottville, NY 14731, USA}
   \date{Received 23 August 2026 / Accepted 24 September 2026}

\abstract{
In recent studies, we have used low-resolution spectroscopy ($R\sim$100)
with the James Webb Space Telescope (JWST) to uncover low-mass 
brown dwarfs (2--10~$M_{\rm Jup}$) in the young cluster IC 348 
that exhibit the so-called 3.4~$\mu$m absorption band from aliphatic
hydrocarbons. Atmospheric models have not predicted the presence of
that band in brown dwarfs at any age. Instead, methane absorption was 
expected in the coolest newborn brown dwarfs, but it was not detected. 
We have performed JWST spectroscopy at higher resolution ($R\sim1000$) on 
one of the brown dwarfs in IC 348, LRL~11001 ($\sim6$~$M_{\rm Jup}$).
The new measurement of the 3.4~$\mu$m band is fit well with 
four Gaussians that represent the CH stretching modes of CH$_2$ and
CH$_3$ groups, indicating that the carrier is a large alkane (C$_n$H$_{2n+2}$). 
In addition, the spectrum contains emission lines at 3.324 and 
3.333~$\mu$m, which we attribute to Q branches of CH$_4$ hot bands between
the octad and dyad regions (primarily $\nu_3+\nu_4\rightarrow\nu_4$) and 
the tetradecad and pentad regions.
This is the first detection of 3.33~$\mu$m CH$_4$ emission from a 
young brown dwarf. The emission probably arises from
the photosphere rather than a circumstellar disk given 
that no disks around stars or brown dwarfs are known to exhibit
emission in these transitions. Modeling of the emission is
needed to constrain the excitation mechanism, which could be either
thermal (e.g., an accretion shock) or pumping by external radiation 
(e.g., neighboring cluster members). 
We speculate that vertical mixing has suppressed CH$_4$ and enhanced large 
alkanes in the photosphere, but CH$_4$ survives in the upper atmosphere,
where it is available to experience the observed emission.}
 
   \keywords{brown dwarfs -- planets and satellites: atmospheres -- stars: atmospheres}

   \maketitle
   \nolinenumbers

\section{Introduction}
\label{sec:intro}

The James Webb Space Telescope \citep[JWST,][]{gar23} has enabled dramatic 
progress in many areas of research, including the study of brown
dwarfs, because of its unprecedented infrared sensitivity, variety of 
observing modes,
and unrestricted access to a wide range of infrared wavelengths. 
In one discovery involving brown dwarfs,
low-resolution spectroscopy with JWST ($R\sim$100) 
identified young low-mass brown dwarfs (2--10~$M_{\rm Jup}$) 
in the star-forming cluster IC~348 
\citep[$\sim$5 Myr, 313~pc, $A_K\lesssim0.4$ mag,][]{luh24ic}
that exhibit absorption in the so-called 3.4~$\mu$m band from non-methane
aliphatic hydrocarbons 
\citep[][hereafter L24 and L25, respectively]{luh24ic,luh25}, 
which were not predicted to appear in brown dwarfs at any age.  Prior to 
that work, the only atmospheres with detections of such hydrocarbons
were those of Saturn and Titan \citep{bel09,dal15}. In a second, seemingly
unrelated discovery, \citet{fah24} detected CH$_4$ emission from 
a binary Y dwarf, CWISEP J193518.59$-$154620.3 
\citep[W1935--15,][]{mar19,def25},
which indicated the presence of a temperature inversion in its atmosphere, 
possibly caused by auroral heating.

To better characterize the hydrocarbon features and other spectral
anomalies among the new brown dwarfs in IC 348, we have performed spectroscopy 
with higher resolution ($R\sim1000$) on the brightest one that
has strong 3.4~$\mu$m absorption, LRL~11001 (source 1 in L24),
which has a mass estimate of $\sim6$~$M_{\rm Jup}$ (L24).
In this Letter, we present our analysis of those data. Most notably, the 
spectrum of LRL~11001 reveals the first detection of 3.33~$\mu$m CH$_4$ 
emission from a young brown dwarf.

\section{JWST/NIRSpec observations}
\label{sec:obs}

We performed spectroscopy on LRL~11001 using the Near-Infrared Spectrograph 
\citep[NIRSpec,][]{jak22} on JWST through
guaranteed time program 4524 (PI: C. Alves de Oliveira) on 2025 August 23 (UT).
The target was observed with the S200A1 fixed slit ($3\farcs2\times0\farcs2$)
and the G235M/F170LP and G395M/F290LP disperser/filter combinations,
covering 1.7--5.1~$\mu$m with a spectral
resolution of $\sim$1000. At each of three nod positions along the slit,
data were collected with the NRSRAPID readout pattern, 320 groups, 
and one integration for a given disperser. The total exposure time was 1500~s
for each disperser. 

To reduce the data for LRL~11001, we applied the JWST Science Calibration 
pipeline version 3.0.0\footnote{\url{https://doi.org/10.5281/zenodo.21364698}}
to the {\tt uncal} files retrieved from the Mikulski Archive for Space 
Telescopes (MAST)\footnote{\url{https://doi.org/10.17909/b9cj-k178}}.
We combined the reduced data from the two dispersers and flux calibrated
the resulting spectrum with NIRCam photometry in F277W, F360M, and F444W (L25).

\begin{figure*}
\includegraphics[width=0.9\textwidth]{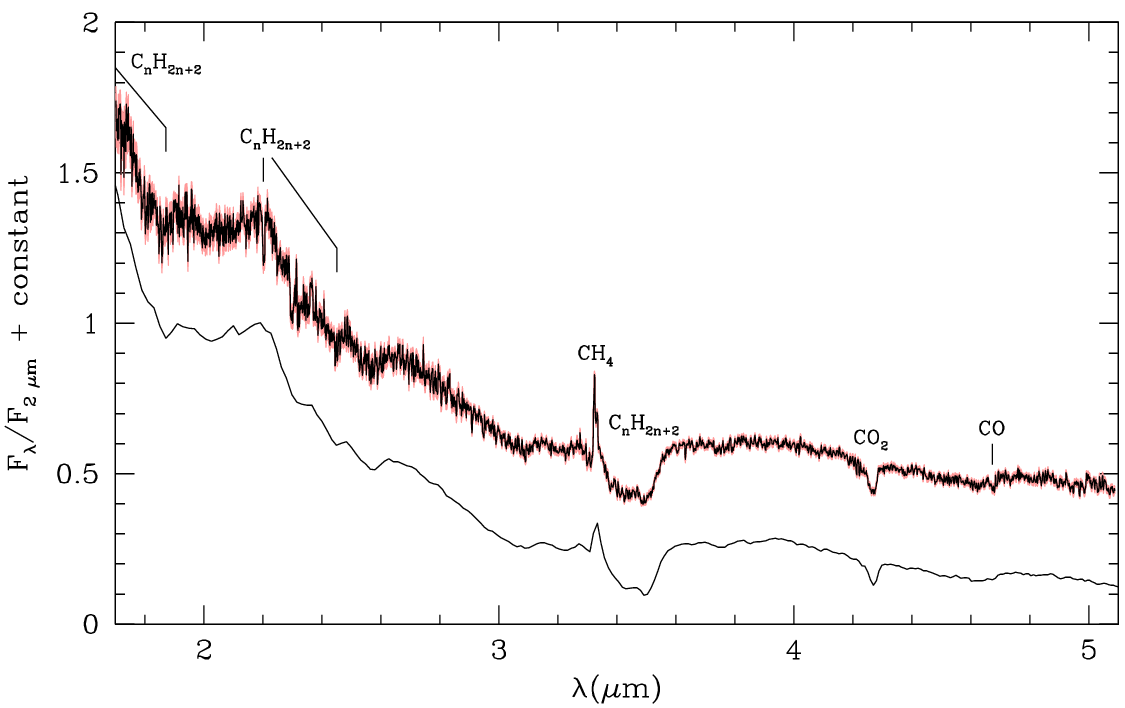}
\caption{JWST/NIRSpec data for LRL~11001 using the PRISM disperser
(bottom, L24) and G235M/G395M (top, this work). Uncertainties
of $\pm1$~$\sigma$ for the latter are plotted as the pink band.
The continuum near the 3.4~$\mu$m band has S/N$\sim$30.
}
\label{fig:spec1}
\end{figure*}

\section{Analysis and results}

\subsection{Absorption features}

We present the NIRSpec G235M/G395M data for LRL~11001 in Fig.~\ref{fig:spec1}.
For comparison, we have included the low-resolution spectrum taken with
the PRISM disperser (L24). The two spectra agree well. We have 
labeled notable features in the new spectrum. H$_2$O bands are present 
across most of the spectrum shortward of 3.2~$\mu$m, which are not labeled.

\begin{figure}
\includegraphics[width=0.95\columnwidth]{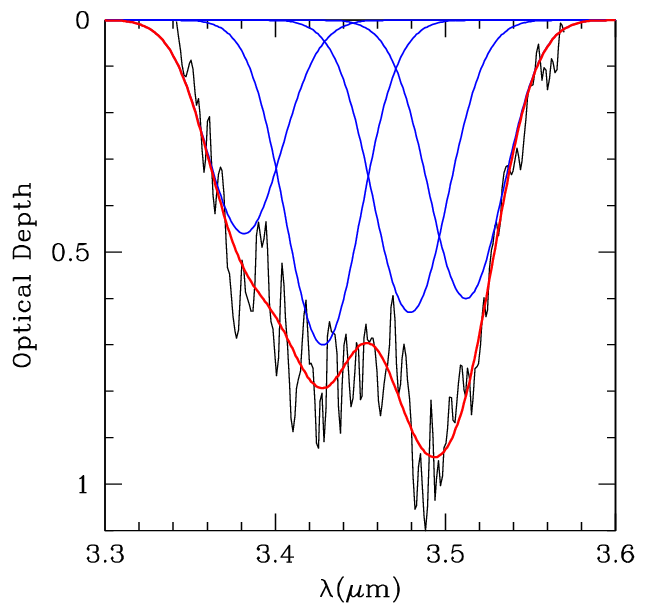}
\caption{Optical depth versus wavelength for the 3.4~$\mu$m band in
the NIRSpec G395M spectrum of LRL~11001. 
We fit the band with four Gaussians that represent the CH stretching 
modes of CH$_2$ and CH$_3$ groups in alkanes \citep{chi13,pen25}.}
\label{fig:spec2}
\end{figure}

In L24 and L25, it was evident from the PRISM data
that the 3.4~$\mu$m absorption detected in LRL~11001 and other IC 348
brown dwarfs closely matched the 3.4~$\mu$m band in the diffuse ISM
in terms of wavelength range \citep{san91,pen94,pen02}.
Our new spectrum at higher resolution for LRL~11001 allows us to characterize
the profile of its 3.4~$\mu$m band. As done in ISM studies \citep{chi13,pen25},
we fit the optical depth of the 3.4~$\mu$m band with four Gaussians
that represent the symmetric and asymmetric CH stretching modes of 
sp$^3$-hybridized bonds in methylene and methyl groups (CH$_2$ and CH$_3$) 
within alkanes (C$_n$H$_{2n+2}$). For the fitting process, we allowed the 
central wavelengths of the Gaussians to vary within $\pm0.006$~$\mu$m from
3.506~$\mu$m (CH$_2$ symmetric), 3.422~$\mu$m (CH$_2$ asymmetric),
3.478~$\mu$m (CH$_3$ symmetric), and 3.381~$\mu$m (CH$_3$ asymmetric),
using the same full width at half maximum for the four Gaussians, which was
allowed to vary between 0.035--0.085~$\mu$m. These ranges of central wavelengths
and widths span the values measured in laboratory spectra \citep{hei98,ris98}.
The amplitudes of the four Gaussians were allowed to vary independently.
The continuum was defined as the average flux at 3.26--3.31 and 
3.57--3.62~$\mu$m.
The best fit Gaussians and their sum are plotted with the G395M data for 
LRL~11001 in Fig.~\ref{fig:spec2}. The good fit of the data with
the four CH stretching modes indicates that the carrier for the
3.4~$\mu$m band has a minimum size of CH$_3$-CH$_2$-CH$_3$ (propane).

Overtone and combination bands from the carrier of the 3.4~$\mu$m band
were detected in the PRISM data for LRL~11001 (L24), and they 
appear in the G235M spectrum as well at 1.7--1.85 and 2.2--2.45~$\mu$m.
In addition, the new data contain a narrow absorption feature at 2.2~$\mu$m
that was unresolved by the PRISM disperser, which is likely part of the 
combination band that begins at 2.2~$\mu$m given that CH$_4$ has a feature 
at a similar wavelength.
Although the fundamental bands of CH$_4$ and larger alkanes differ 
significantly in wavelengths, low-resolution laboratory spectra suggest that 
their overtone and combination bands are more closely aligned \citep{cla09}.

Both the PRISM and G395M spectra show absorption in CO$_2$ at 4.27~$\mu$m.
The G395M data also contains weak absorption from CO at 4.67~$\mu$m.
The narrow widths of the two features are inconsistent with photospheric
molecules \citep{yam10,bei23b} and instead are indicative of ices 
\citep{boo15,mcc23b}, which could reside in the circumstellar environment of 
the brown dwarf or the surrounding molecular cloud.

\subsection{Methane emission}

The PRISM spectrum of LRL~11001 exhibits a peak at 3.33~$\mu$m, which we 
originally interpreted as the continuum between the 3.4~$\mu$m band and broad
absorption at shorter wavelengths, perhaps from H$_2$O (photospheric or 
interstellar). However, the higher spectral resolution of the G395M data 
reveals that the 3.33~$\mu$m peak is an emission feature 
(Fig.~\ref{fig:spec1}). In the background-subtracted spectral images for 
each of the three nods, the emission is unresolved and coincides with 
LRL~11001.

\begin{figure}
\centering
\includegraphics[width=0.95\columnwidth]{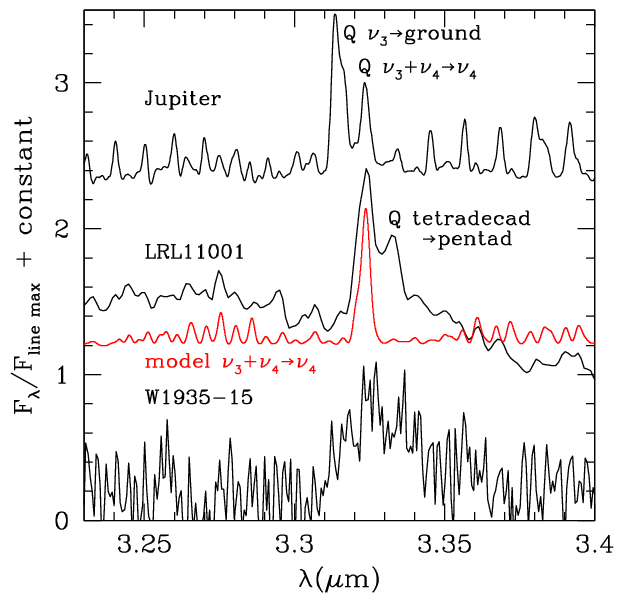}
\caption{Spectra of Jupiter \citep[ISO SWS,][]{enc96,enc99}, LRL~11001
(NIRSpec G395M, this work), the Y dwarf W1935--15 
\citep[NIRSpec G395H,][]{fah24},
and a model of the $\nu_3+\nu_4\rightarrow\nu_4$ band of CH$_4$
\citep{san22}. In the spectrum of Jupiter, the two brightest 
lines correspond to the Q branches for $\nu_3\rightarrow$ground 
and hot bands between the octad and dyad regions (primarily 
$\nu_3+\nu_4\rightarrow\nu_4$). They are surrounded by the R and P branches of 
those bands. We identify the brightest pair of emission lines in LRL~11001 as 
the Q branches for transitions between the octad and dyad regions and between 
the tetradecad and pentad regions.
The surrounding spectrum of LRL~11001 may contain weak detections
the R and P branch lines for $\nu_3+\nu_4\rightarrow\nu_4$, as illustrated
by the comparison to the model spectrum.
The emission for W1935--15 is aligned with the pair of lines in LRL~11001,
although the former may be broader, possibly including emission from the 
Q branch of $\nu_3\rightarrow$ground.}
\label{fig:spec3}
\end{figure}

We have identified CH$_4$ as the carrier for the 3.33~$\mu$m emission
from LRL~11001. In Fig.~\ref{fig:spec3}, we compare the G395M data in the 
vicinity of the emission to spectra for Jupiter and W1935--15.
The spectrum of Jupiter was obtained by \citet{enc96} using the
Short Wavelength Spectrometer \citep[SWS,][]{deg96} on the Infrared Space
Observatory \citep[ISO,][]{kes96}. Those data contain fluorescent CH$_4$
emission that is driven by solar radiation \citep{dro99,enc99}.
The emission lines in Jupiter's spectrum include the Q branch for 
$\nu_3\rightarrow$ground (3.314~$\mu$m), the Q branches for hot bands between
the octad and dyad regions (primarily $\nu_3+\nu_4\rightarrow\nu_4$;
3.324~$\mu$m), and the accompanying P and R branches on either side of the 
Q branches. 

The emission feature in LRL~11001 consists of a pair of partially blended 
lines, as shown in Fig.~\ref{fig:spec3}.
The brighter line (3.324~$\mu$m) coincides with the Q branch for 
$\nu_3+\nu_4\rightarrow\nu_4$. We attribute the fainter line (3.333~$\mu$m)
to the Q branches of hot bands between the tetradecad and pentad regions 
(e.g., 2$\nu_3\rightarrow\nu_3$). The P and R branches for these bands
should also be produced at some level. To serve as a guide for the wavelengths
expected for the P and R branches for the $\nu_3+\nu_4\rightarrow\nu_4$ band,
we have included in Fig.~\ref{fig:spec3} a model spectrum of the P, Q, and 
R branches of the $\nu_3+\nu_4\rightarrow\nu_4$ band for Jupiter \citep{san22}.
The spectrum of LRL~11001 may contain weak detections of some of those
P and R branch lines. The P branch lines at $>$3.37~$\mu$m may be masked 
by the absorption in the 3.4~$\mu$m band. The Q branch for 
$\nu_3\rightarrow$ground is not detected in the spectrum of LRL~11001. 

The two emission lines in LRL~11001 are aligned with the emission feature in 
the Y dwarf W1935--15, although the latter is broader and may include 
emission from the Q branch of $\nu_3\rightarrow$ground. The differences in 
the strengths of that branch relative to other Q branches among Jupiter, 
W1935--15, and LRL~11001 may reflect differences in effective 
temperatures \citep[126, 482$\pm$38, $\sim$1500--1600~K,][L24]{li12,fah24}
as well as differences in excitation mechanisms.

\section{Discussion and conclusions}

We have presented medium-resolution ($\sim1000$) spectroscopy at
1.7--5.1~$\mu$m for LRL~11001, which is a $\sim6$~$M_{\rm Jup}$ brown
dwarf in IC 348 that exhibited a strong 3.4~$\mu$m band in its previous
low-resolution spectrum. Most notably, the new data reveal the presence
of a pair of emission lines near 3.33~$\mu$m, which we attribute to the
Q branches of CH$_4$ for transitions between the octad and dyad regions 
and between the tetradecad and pentad regions. This is the first
detection of 3.33~$\mu$m CH$_4$ emission from a young brown dwarf.

Previous studies have detected weak emission in the deformation 
$\nu_4$ mode of CH$_4$ at 7.65~$\mu$m from young stars and brown dwarfs, 
which has been attributed to their disks 
\citep{tab23,ara25,fla25,kan26}. The 3.33~$\mu$m emission in LRL~11001 
is incompatible with gas of that kind given the very different level
populations implied by the emitting bands. The upper levels for the 
7.65~$\mu$m transitions have excitation energies of 1300~cm$^{-1}$, but 
for LRL~11001 the $\nu_3\rightarrow$ground band is absent (3000~cm$^{-1}$) 
and only the hot bands are observed (4500--6000~cm$^{-1}$).

In general, line emission from a stellar source can arise from either
thermally excited gas in an atmosphere that has a temperature inversion
or from gas that is pumped by external radiation (fluorescent emission).
For instance, the CH$_4$ lines from Jupiter were originally modeled
as thermal emission \citep{enc96}, but were later found to fit better
with fluorescence that is driven by solar radiation \citep{dro99,enc99}.
Only the former scenario has been considered for the binary Y dwarf W1935--15
since a source of irradiation appears to be absent \citep{fah24,sua25,smi26}.
The fluorescent emission from Jupiter and the thermal emission
from W1935--15 originate from $\mu$bar and mbar pressure levels, respectively
\citep{fah24,san22}, so the CH$_4$ emission from LRL~11001 may be produced by
its upper atmosphere as well.
Modeling of the CH$_4$ emission from LRL~11001 is needed to constrain
the excitation mechanism and the location of the emitting gas. 
An accretion shock from a circumstellar disk
is one option for the deposition of heat into the atmosphere.
The 1--5~$\mu$m data from NIRSpec do not show excess emission from a disk
relative to other brown dwarfs in IC~348,
but an excess could be present at longer wavelengths where no data are 
available. Possible sources of irradiation include an accretion shock and 
neighboring stars in IC~348.
In theory, the emitting CH$_4$ for these scenarios could reside in either 
the brown dwarf's atmosphere or the surface of its disk (if it has one), 
but the former seems more likely given the apparent connection between 
the CH$_4$ emission and the 3.4~$\mu$m band and the fact that the latter
is produced by the brown dwarf rather than a disk (e.g., the stellar 
photosphere dominates throughout wavelength range of the hydrocarbon features).
In addition, if an irradiated disk around LRL~11001 is the source of 
the 3.33~$\mu$m CH$_4$ emission, it is unclear why such emission has never been
observed from disks around other young stars and brown dwarfs.

The excitation mechanism for the CH$_4$ emission could be constrained
through medium-resolution spectroscopy of the remaining brown dwarfs
in IC 348 that show the 3.4~$\mu$m band and similar objects in other
regions as they are found. Such data could be used to search for
correlations between CH$_4$ emission and radiation environment and the
presence of disks. Spectroscopy longward of 5~$\mu$m is also needed for
these brown dwarfs to check for excess emission from disks.

We also discuss the origin of the 3.4~$\mu$m band, and new insight provided
by the detection of CH$_4$ emission.
CH$_4$ absorption at 3.33~$\mu$m is weaker at younger ages among L dwarfs
\citep{mil18,mil23,bei23a} and is absent among the coolest known brown dwarfs
at ages of $\leq$~10~Myr \citep[][L24,L25]{luh23}, which may be caused
by enhanced disequilibrium chemistry from vertical mixing \citep{for20}. 
On the other hand, invoking disequilibrium chemistry may be unnecessary 
to explain some nondetections of CH$_4$ in young L dwarfs \citep{zha25}.
However, the combined absence of CH$_4$ and presence of the alkane carriers
for the 3.4~$\mu$m band in the coolest IC 348 brown dwarfs may 
represent direct evidence of disequilibrium chemistry. 
One mechanism for disequilibrium chemistry is vertical mixing, which is
expected to affect the abundances of hydrocarbons \citep{mos11,bil13}.
We speculate that the absence of CH$_4$ and the presence of larger alkanes 
in the photosphere result from vertical mixing, and that CH$_4$
survives in the upper atmosphere, where it is available to experience the 
emission observed in LRL~11001.
Another option is external irradiation, which could lead to the destruction
of CH$_4$ and the formation of more complex hydrocarbons 
\citep{mos16,zah16,hel20}. 
However, irradiation would not explain the trend of weaker CH$_4$ at 
younger ages that is observed among all L dwarfs. Ideally, we would prefer 
a single model that explains the trend of CH$_4$ with age, 
the presence of the carrier of the 3.4~$\mu$m band among the coolest newborn
brown dwarfs, the temporary re-emergence of TiO and VO bands and weakening
of H$_2$O bands after the onset of 3.4~$\mu$m absorption (L25),
and the presence of the emitting CH$_4$ in LRL~11001.

\section*{Data availability}

The spectrum from Sect.~\ref{sec:obs} is available in electronic form at CDS.

\begin{acknowledgements}

We thank the referee, Genaro Su{\'a}rez, for their careful and constructive 
report. We thank Pierre Drossart, Tom Geballe, Emmanuel Lellouch, Yvonne 
Pendleton, Ewine van Dishoeck, Channon Visscher, and Ger{\'o}nimo Villanueva 
for helpful discussions. 
This work is based on observations made with the NASA/ESA/CSA James Webb Space 
Telescope. The data are associated with program 4524 and were obtained from
MAST at the Space Telescope Science Institute, which is operated by the 
Association of Universities for Research in Astronomy, Inc., under NASA 
contract NAS 5-03127. The Center for Exoplanets and Habitable Worlds is 
supported by the Pennsylvania State University, the Eberly College of Science, 
and the Pennsylvania Space Grant Consortium.

\end{acknowledgements}

\bibliographystyle{aa} 
\bibliography{ref} 

@ARTICLE{gar23,
       author = {{Gardner}, Jonathan P. and {Mather}, John C. and {Abbott}, Randy and {Abell}, James S. and {Abernathy}, Mark and {Abney}, Faith E. and {Abraham}, John G. and {Abraham}, Roberto and {Abul-Huda}, Yasin M. and {Acton}, Scott and {Adams}, Cynthia K. and {Adams}, Evan and {Adler}, David S. and {Adriaensen}, Maarten and {Aguilar}, Jonathan Albert and {Ahmed}, Mansoor and {Ahmed}, Nasif S. and {Ahmed}, Tanjira and {Albat}, R{\"u}deger and {Albert}, Lo{\"\i}c and {Alberts}, Stacey and {Aldridge}, David and {Allen}, Mary Marsha and {Allen}, Shaune S. and {Altenburg}, Martin and {Altunc}, Serhat and {Alvarez}, Jose Lorenzo and {{\'A}lvarez-M{\'a}rquez}, Javier and {Alves de Oliveira}, Catarina and {Ambrose}, Leslie L. and {Anandakrishnan}, Satya M. and {Andersen}, Gregory C. and {Anderson}, Harry James and {Anderson}, Jay and {Anderson}, Kristen and {Anderson}, Sara M. and {Aprea}, Julio and {Archer}, Benita J. and {Arenberg}, Jonathan W. and {Argyriou}, Ioannis and {Arribas}, Santiago and {Artigau}, {\'E}tienne and {Arvai}, Amanda Rose and {Atcheson}, Paul and {Atkinson}, Charles B. and {Averbukh}, Jesse and {Aymergen}, Cagatay and {Bacinski}, John J. and {Baggett}, Wayne E. and {Bagnasco}, Giorgio and {Baker}, Lynn L. and {Balzano}, Vicki Ann and {Banks}, Kimberly A. and {Baran}, David A. and {Barker}, Elizabeth A. and {Barrett}, Larry K. and {Barringer}, Bruce O. and {Barto}, Allison and {Bast}, William and {Baudoz}, Pierre and {Baum}, Stefi and {Beatty}, Thomas G. and {Beaulieu}, Mathilde and {Bechtold}, Kathryn and {Beck}, Tracy and {Beddard}, Megan M. and {Beichman}, Charles and {Bellagama}, Larry and {Bely}, Pierre and {Berger}, Timothy W. and {Bergeron}, Louis E. and {Bernier}, Antoine-Darveau and {Bertch}, Maria D. and {Beskow}, Charlotte and {Betz}, Laura E. and {Biagetti}, Carl P. and {Birkmann}, Stephan and {Bjorklund}, Kurt F. and {Blackwood}, James D. and {Blazek}, Ronald Paul and {Blossfeld}, Stephen and {Bluth}, Marcel and {Boccaletti}, Anthony and {Boegner}, Jr., Martin E. and {Bohlin}, Ralph C. and {Boia}, John Joseph and {B{\"o}ker}, Torsten and {Bonaventura}, N. and {Bond}, Nicholas A. and {Bosley}, Kari Ann and {Boucarut}, Rene A. and {Bouchet}, Patrice and {Bouwman}, Jeroen and {Bower}, Gary and {Bowers}, Ariel S. and {Bowers}, Charles W. and {Boyce}, Leslye A. and {Boyer}, Christine T. and {Boyer}, Martha L. and {Boyer}, Michael and {Boyer}, Robert and {Bradley}, Larry D. and {Brady}, Gregory R. and {Brandl}, Bernhard R. and {Brannen}, Judith L. and {Breda}, David and {Bremmer}, Harold G. and {Brennan}, David and {Bresnahan}, Pamela A. and {Bright}, Stacey N. and {Broiles}, Brian J. and {Bromenschenkel}, Asa and {Brooks}, Brian H. and {Brooks}, Keira J. and {Brown}, Bob and {Brown}, Bruce and {Brown}, Thomas M. and {Bruce}, Barry W. and {Bryson}, Jonathan G. and {Bujanda}, Edwin D. and {Bullock}, Blake M. and {Bunker}, A.~J. and {Bureo}, Rafael and {Burt}, Irving J. and {Bush}, James Aaron and {Bushouse}, Howard A. and {Bussman}, Marie C. and {Cabaud}, Olivier and {Cale}, Steven and {Calhoon}, Charles D. and {Calvani}, Humberto and {Canipe}, Alicia M. and {Caputo}, Francis M. and {Cara}, Mihai and {Carey}, Larkin and {Case}, Michael Eli and {Cesari}, Thaddeus and {Cetorelli}, Lee D. and {Chance}, Don R. and {Chandler}, Lynn and {Chaney}, Dave and {Chapman}, George N. and {Charlot}, S. and {Chayer}, Pierre and {Cheezum}, Jeffrey I. and {Chen}, Bin and {Chen}, Christine H. and {Cherinka}, Brian and {Chichester}, Sarah C. and {Chilton}, Zachary S. and {Chittiraibalan}, Dharini and {Clampin}, Mark and {Clark}, Charles R. and {Clark}, Kerry W. and {Clark}, Stephanie M. and {Claybrooks}, Edward E. and {Cleveland}, Keith A. and {Cohen}, Andrew L. and {Cohen}, Lester M. and {Col{\'o}n}, Knicole D. and {Coleman}, Benee L. and {Colina}, Luis and {Comber}, Brian J. and {Comeau}, Thomas M. and {Comer}, Thomas and {Conde Reis}, Alain and {Connolly}, Dennis C. and {Conroy}, Kyle E. and {Contos}, Adam R. and {Contreras}, James and {Cook}, Neil J. and {Cooper}, James L. and {Cooper}, Rachel Aviva and {Correia}, Michael F. and {Correnti}, Matteo and {Cossou}, Christophe and {Costanza}, Brian F. and {Coulais}, Alain and {Cox}, Colin R. and {Coyle}, Ray T. and {Cracraft}, Misty M. and {Crew}, Keith A. and {Curtis}, Gary J. and {Cusveller}, Bianca and {Da Costa Maciel}, Cleyciane and {Dailey}, Christopher T. and {Daugeron}, Fr{\'e}d{\'e}ric and {Davidson}, Greg S. and {Davies}, James E. and {Davis}, Katherine Anne and {Davis}, Michael S. and {Day}, Ratna and {de Chambure}, Daniel and {de Jong}, Pauline and {De Marchi}, Guido and {Dean}, Bruce H. and {Decker}, John E. and {Delisa}, Amy S. and {Dell}, Lawrence C. and {Dellagatta}, Gail},
        title = "{The James Webb Space Telescope Mission}",
      journal = {\pasp},
         year = 2023,
        month = jun,
       volume = {135},
       number = {1048},
          eid = {068001},
        pages = {068001},
          doi = {10.1088/1538-3873/acd1b5},
archivePrefix = {arXiv},
       eprint = {2304.04869},
 primaryClass = {astro-ph.IM},
       adsurl = {https://ui.adsabs.harvard.edu/abs/2023PASP..135f8001G}
}

@ARTICLE{jak22,
       author = {{Jakobsen}, P. and {Ferruit}, P. and {Alves de Oliveira}, C. and {Arribas}, S. and {Bagnasco}, G. and {Barho}, R. and {Beck}, T.~L. and {Birkmann}, S. and {B{\"o}ker}, T. and {Bunker}, A.~J. and {Charlot}, S. and {de Jong}, P. and {de Marchi}, G. and {Ehrenwinkler}, R. and {Falcolini}, M. and {Fels}, R. and {Franx}, M. and {Franz}, D. and {Funke}, M. and {Giardino}, G. and {Gnata}, X. and {Holota}, W. and {Honnen}, K. and {Jensen}, P.~L. and {Jentsch}, M. and {Johnson}, T. and {Jollet}, D. and {Karl}, H. and {Kling}, G. and {K{\"o}hler}, J. and {Kolm}, M. -G. and {Kumari}, N. and {Lander}, M.~E. and {Lemke}, R. and {L{\'o}pez-Caniego}, M. and {L{\"u}tzgendorf}, N. and {Maiolino}, R. and {Manjavacas}, E. and {Marston}, A. and {Maschmann}, M. and {Maurer}, R. and {Messerschmidt}, B. and {Moseley}, S.~H. and {Mosner}, P. and {Mott}, D.~B. and {Muzerolle}, J. and {Pirzkal}, N. and {Pittet}, J. -F. and {Plitzke}, A. and {Posselt}, W. and {Rapp}, B. and {Rauscher}, B.~J. and {Rawle}, T. and {Rix}, H. -W. and {R{\"o}del}, A. and {Rumler}, P. and {Sabbi}, E. and {Salvignol}, J. -C. and {Schmid}, T. and {Sirianni}, M. and {Smith}, C. and {Strada}, P. and {te Plate}, M. and {Valenti}, J. and {Wettemann}, T. and {Wiehe}, T. and {Wiesmayer}, M. and {Willott}, C.~J. and {Wright}, R. and {Zeidler}, P. and {Zincke}, C.},
        title = "{The Near-Infrared Spectrograph (NIRSpec) on the James Webb Space Telescope. I. Overview of the instrument and its capabilities}",
      journal = {\aap},
         year = 2022,
        month = may,
       volume = {661},
          eid = {A80},
        pages = {A80},
          doi = {10.1051/0004-6361/202142663},
archivePrefix = {arXiv},
       eprint = {2202.03305},
 primaryClass = {astro-ph.IM},
       adsurl = {https://ui.adsabs.harvard.edu/abs/2022A&A...661A..80J}
}

@ARTICLE{luh25,
       author = {{Luhman}, K.~L. and {Alves de Oliveira}, C.},
        title = "{A New Spectral Class of Brown Dwarfs at the Bottom of the IMF in IC 348}",
      journal = {\apjl},
         year = 2025,
        month = jun,
       volume = {986},
       number = {1},
          eid = {L14},
        pages = {L14},
          doi = {10.3847/2041-8213/addc55},
       adsurl = {https://ui.adsabs.harvard.edu/abs/2025ApJ...986L..14L}
}

@ARTICLE{luh24ic,
       author = {{Luhman}, K.~L. and {Alves de Oliveira}, C. and {Baraffe}, I. and {Chabrier}, G. and {Geballe}, T.~R. and {Parker}, R.~J. and {Pendleton}, Y.~J. and {Tremblin}, P.},
        title = "{A JWST Survey for Planetary Mass Brown Dwarfs in IC 348}",
      journal = {\aj},
         year = 2024,
        month = jan,
       volume = {167},
       number = {1},
          eid = {19},
        pages = {19},
          doi = {10.3847/1538-3881/ad00b7},
       adsurl = {https://ui.adsabs.harvard.edu/abs/2024AJ....167...19L}
}

@ARTICLE{luh23,
       author = {{Luhman}, K.~L. and {Tremblin}, P. and {Birkmann}, S.~M. and {Manjavacas}, E. and {Valenti}, J. and {Alves de Oliveira}, C. and {Beck}, T.~L. and {Giardino}, G. and {L{\"u}tzgendorf}, N. and {Rauscher}, B.~J. and {Sirianni}, M.},
        title = "{JWST/NIRSpec Observations of the Planetary Mass Companion TWA 27B}",
      journal = {\apjl},
         year = 2023,
        month = jun,
       volume = {949},
       number = {2},
          eid = {L36},
        pages = {L36},
          doi = {10.3847/2041-8213/acd635},
archivePrefix = {arXiv},
       eprint = {2305.18603},
 primaryClass = {astro-ph.EP},
       adsurl = {https://ui.adsabs.harvard.edu/abs/2023ApJ...949L..36L}
}

@ARTICLE{def25,
       author = {{De Furio}, Matthew and {Meyer}, Michael R. and {Greene}, Thomas and {Hodapp}, Klaus and {Johnstone}, Doug and {Leisenring}, Jarron and {Rieke}, Marcia and {Robberto}, Massimo and {Roellig}, Thomas and {Cugno}, Gabriele and {Fiorellino}, Eleonora and {Manara}, Carlo F. and {Raileanu}, Roberta and {van Terwisga}, Sierk},
        title = "{Identification of a Turnover in the Initial Mass Function of a Young Stellar Cluster Down to 0.5 M$_{J}$}",
      journal = {\apjl},
         year = 2025,
        month = mar,
       volume = {981},
       number = {2},
          eid = {L34},
        pages = {L34},
          doi = {10.3847/2041-8213/adb96a},
archivePrefix = {arXiv},
       eprint = {2409.04624},
 primaryClass = {astro-ph.SR},
       adsurl = {https://ui.adsabs.harvard.edu/abs/2025ApJ...981L..34D}
}

@ARTICLE{fah24,
       author = {{Faherty}, Jacqueline K. and {Burningham}, Ben and {Gagn{\'e}}, Jonathan and {Su{\'a}rez}, Genaro and {Vos}, Johanna M. and {Alejandro Merchan}, Sherelyn and {Morley}, Caroline V. and {Rowland}, Melanie and {Lacy}, Brianna and {Kiman}, Rocio and {Caselden}, Dan and {Kirkpatrick}, J. Davy and {Meisner}, Aaron and {Schneider}, Adam C. and {Kuchner}, Marc Jason and {Bardalez Gagliuffi}, Daniella Carolina and {Beichman}, Charles and {Eisenhardt}, Peter and {Gelino}, Christopher R. and {Gharib-Nezhad}, Ehsan and {Gonzales}, Eileen and {Marocco}, Federico and {Rothermich}, Austin James and {Whiteford}, Niall},
        title = "{Methane emission from a cool brown dwarf}",
      journal = {\nat},
         year = 2024,
        month = apr,
       volume = {628},
       number = {8008},
        pages = {511-514},
          doi = {10.1038/s41586-024-07190-w},
archivePrefix = {arXiv},
       eprint = {2404.10977},
 primaryClass = {astro-ph.SR},
       adsurl = {https://ui.adsabs.harvard.edu/abs/2024Natur.628..511F}
}

@ARTICLE{ris98,
       author = {{Ristein}, J. and {Stief}, R. T. and {Beyer}, W.},
        title = "{A comparative analysis of a-C:H by infrared spectroscopy and mass selected thermal effusion }",
      journal = {JAP},
         year = 1998,
        month = oct,
       volume = {84},
          eid = {3836},
        pages = {3836},
          doi = {10.1063/1.368563}
}

@ARTICLE{sua25,
       author = {{Su{\'a}rez}, Genaro and {Faherty}, Jacqueline K. and {Burningham}, Ben and {Morley}, Caroline V. and {Vos}, Johanna M. and {Lacy}, Brianna and {Rowland}, Melanie J. and {Schneider}, Adam C. and {Alejandro Merchan}, Sherelyn and {Bardalez Gagliuffi}, Daniella C. and {Bickle}, Thomas P. and {Gonzales}, Eileen C. and {Kiman}, Rocio and {Rothermich}, Austin and {Whiteford}, Niall},
        title = "{Diversity of Cold Worlds: Predicted Near-to-mid-infrared Spectral Signatures of a Cold Brown Dwarf with Potential Auroral Heating}",
      journal = {\apj},
         year = 2025,
        month = nov,
       volume = {993},
       number = {2},
          eid = {165},
        pages = {165},
          doi = {10.3847/1538-4357/ae0e6a},
archivePrefix = {arXiv},
       eprint = {2509.26505},
 primaryClass = {astro-ph.SR},
       adsurl = {https://ui.adsabs.harvard.edu/abs/2025ApJ...993..165S}
}

@ARTICLE{pen25,
       author = {{Pendleton}, Yvonne J. and {Geballe}, T.~R. and {Chu}, Laurie E.~U. and {Decleir}, Marjorie and {Gordon}, Karl D. and {Tielens}, A.~G.~G.~M. and {Allamandola}, Louis J. and {Bouwman}, Jeroen and {Chiar}, J.~E. and {Dewitt}, Curtis and {Gunay}, Burcu and {Henning}, Thomas and {Mennella}, Vito and {Palumbo}, M.~E. and {Potapov}, Alexey and {Rashman}, Maisie and {Zeegers}, Sascha},
        title = "{A Tale of Two Sightlines: Comparison of Hydrocarbon Dust Absorption Bands toward Cygnus OB2-12 and the Galactic Center}",
      journal = {\apj},
         year = 2025,
        month = oct,
       volume = {992},
       number = {1},
          eid = {8},
        pages = {8},
          doi = {10.3847/1538-4357/adfc3d},
archivePrefix = {arXiv},
       eprint = {2508.12601},
 primaryClass = {astro-ph.GA},
       adsurl = {https://ui.adsabs.harvard.edu/abs/2025ApJ...992....8P}
}

@ARTICLE{chi13,
       author = {{Chiar}, J.~E. and {Tielens}, A.~G.~G.~M. and {Adamson}, A.~J. and {Ricca}, A.},
        title = "{The Structure, Origin, and Evolution of Interstellar Hydrocarbon Grains}",
      journal = {\apj},
         year = 2013,
        month = jun,
       volume = {770},
       number = {1},
          eid = {78},
        pages = {78},
          doi = {10.1088/0004-637X/770/1/78},
       adsurl = {https://ui.adsabs.harvard.edu/abs/2013ApJ...770...78C}
}

@ARTICLE{enc96,
       author = {{Encrenaz}, T. and {de Graauw}, T. and {Schaeidt}, S. and {Lellouch}, E. and {Feuchtgruber}, H. and {Beintema}, D.~A. and {Bezard}, B. and {Drossart}, P. and {Griffin}, M. and {Heras}, A. and {Kessler}, M. and {Leech}, K. and {Morris}, P. and {Roelfsema}, P.~R. and {Roos-Serote}, M. and {Salama}, A. and {Vandenbussche}, B. and {Valentijn}, E.~A. and {Davis}, G.~R. and {Naylor}, D.~A.},
        title = "{First results of ISO-SWS observations of Jupiter.}",
      journal = {\aap},
         year = 1996,
        month = nov,
       volume = {315},
        pages = {L397-L400},
       adsurl = {https://ui.adsabs.harvard.edu/abs/1996A&A...315L.397E}
}

@ARTICLE{enc99,
       author = {{Encrenaz}, Th. and {Drossart}, P. and {Feuchtgruber}, H. and {Lellouch}, E. and {B{\'e}zard}, B. and {Fouchet}, T. and {Atreya}, S.~K.},
        title = "{The atmospheric composition and structure of Jupiter and Saturn from ISO observations: a preliminary review}",
      journal = {\planss},
         year = 1999,
        month = oct,
       volume = {47},
       number = {10-11},
        pages = {1225-1242},
          doi = {10.1016/S0032-0633(99)00046-X},
       adsurl = {https://ui.adsabs.harvard.edu/abs/1999P&SS...47.1225E}
}

@ARTICLE{hei98,
       author = {{Heitz}, T. and {Dr{\'e}villon}, B. and {Godet}, C. and {Bour{\'e}e}, J.~E.},
        title = "{Quantitative study of C-H bonding in polymerlike amorphous carbon films using in situ infrared ellipsometry}",
      journal = {\prb},
         year = 1998,
        month = nov,
       volume = {58},
       number = {20},
        pages = {13957-13973},
          doi = {10.1103/PhysRevB.58.13957},
       adsurl = {https://ui.adsabs.harvard.edu/abs/1998PhRvB..5813957H}
}

@ARTICLE{san22,
       author = {{S{\'a}nchez-L{\'o}pez}, A. and {L{\'o}pez-Puertas}, M. and {Garc{\'\i}a-Comas}, M. and {Funke}, B. and {Fouchet}, T. and {Snellen}, I.~A.~G.},
        title = "{The CH$_{4}$ abundance in Jupiter's upper atmosphere}",
      journal = {\aap},
         year = 2022,
        month = jun,
       volume = {662},
          eid = {A91},
        pages = {A91},
          doi = {10.1051/0004-6361/202141933},
archivePrefix = {arXiv},
       eprint = {2203.10086},
 primaryClass = {astro-ph.EP},
       adsurl = {https://ui.adsabs.harvard.edu/abs/2022A&A...662A..91S}
}

@ARTICLE{tab23,
       author = {{Tabone}, B. and {Bettoni}, G. and {van Dishoeck}, E.~F. and {Arabhavi}, A.~M. and {Grant}, S. and {Gasman}, D. and {Henning}, Th. and {Kamp}, I. and {G{\"u}del}, M. and {Lagage}, P.~O. and {Ray}, T. and {Vandenbussche}, B. and {Abergel}, A. and {Absil}, O. and {Argyriou}, I. and {Barrado}, D. and {Boccaletti}, A. and {Bouwman}, J. and {Caratti o Garatti}, A. and {Geers}, V. and {Glauser}, A.~M. and {Justannont}, K. and {Lahuis}, F. and {Mueller}, M. and {Nehm{\'e}}, C. and {Olofsson}, G. and {Pantin}, E. and {Scheithauer}, S. and {Waelkens}, C. and {Waters}, L.~B.~F.~M. and {Black}, J.~H. and {Christiaens}, V. and {Guadarrama}, R. and {Morales-Calder{\'o}n}, M. and {Jang}, H. and {Kanwar}, J. and {Pawellek}, N. and {Perotti}, G. and {Perrin}, A. and {Rodgers-Lee}, D. and {Samland}, M. and {Schreiber}, J. and {Schwarz}, K. and {Colina}, L. and {{\"O}stlin}, G. and {Wright}, G.},
        title = "{A rich hydrocarbon chemistry and high C to O ratio in the inner disk around a very low-mass star}",
      journal = {Nature Astronomy},
         year = 2023,
        month = jul,
       volume = {7},
        pages = {805-814},
          doi = {10.1038/s41550-023-01965-3},
archivePrefix = {arXiv},
       eprint = {2304.05954},
 primaryClass = {astro-ph.EP},
       adsurl = {https://ui.adsabs.harvard.edu/abs/2023NatAs...7..805T}
}

@ARTICLE{kan26,
       author = {{Kanwar}, Jayatee and {Kamp}, Inga and {Woitke}, Peter and {van Dishoeck}, Ewine F. and {Henning}, Thomas and {Liu}, Yao and {Kaeufer}, Till and {Tabone}, Beno{\^\i}t and {G{\"u}del}, Manuel and {Barrado}, David and {Arabhavi}, Aditya M. and {Franceschi}, Riccardo and {Vlasblom}, Marissa},
        title = "{MINDS: Strong oxygen depletion in the inner regions of a very low-mass star disk?}",
      journal = {\aap},
         year = 2026,
        month = jan,
       volume = {705},
          eid = {A222},
        pages = {A222},
          doi = {10.1051/0004-6361/202451844},
archivePrefix = {arXiv},
       eprint = {2508.11761},
 primaryClass = {astro-ph.EP},
       adsurl = {https://ui.adsabs.harvard.edu/abs/2026A&A...705A.222K}
}

@ARTICLE{ara25,
       author = {{Arabhavi}, Aditya M. and {Kamp}, Inga and {van Dishoeck}, Ewine F. and {Henning}, Thomas and {Jang}, Hyerin and {Christiaens}, Valentin and {Gasman}, Danny and {Pascucci}, Ilaria and {Perotti}, Giulia and {Grant}, Sierra L. and {Barrado}, David and {G{\"u}del}, Manuel and {Lagage}, Pierre-Olivier and {Caratti o Garatti}, Alessio and {Lahuis}, Fred and {Waters}, L.~B.~F.~M. and {Kaeufer}, Till and {Kanwar}, Jayatee and {Morales-Calder{\'o}n}, Maria and {Schwarz}, Kamber and {Sellek}, Andrew D. and {Tabone}, Beno{\^\i}t and {Temmink}, Milou and {Vlasblom}, Marissa},
        title = "{MINDS: The Very Low-mass Star and Brown Dwarf Sample Hidden Water in Carbon-dominated Protoplanetary Disks}",
      journal = {\apjl},
         year = 2025,
        month = may,
       volume = {984},
       number = {2},
          eid = {L62},
        pages = {L62},
          doi = {10.3847/2041-8213/adc692},
archivePrefix = {arXiv},
       eprint = {2504.11425},
 primaryClass = {astro-ph.EP},
       adsurl = {https://ui.adsabs.harvard.edu/abs/2025ApJ...984L..62A}
}

@ARTICLE{fla25,
       author = {{Flagg}, Laura and {Scholz}, Aleks and {Almendros-Abad}, V. and {Jayawardhana}, Ray and {Damian}, Belinda and {Mu{\v{z}}i{\'c}}, Koraljka and {Natta}, Antonella and {Pinilla}, Paola and {Testi}, Leonardo},
        title = "{Detection of Hydrocarbons in the Disk around an Actively Accreting Planetary-mass Object}",
      journal = {\apj},
         year = 2025,
        month = jun,
       volume = {986},
       number = {2},
          eid = {200},
        pages = {200},
          doi = {10.3847/1538-4357/add71d},
archivePrefix = {arXiv},
       eprint = {2505.13714},
 primaryClass = {astro-ph.EP},
       adsurl = {https://ui.adsabs.harvard.edu/abs/2025ApJ...986..200F}
}

@ARTICLE{pen94,
       author = {{Pendleton}, Y.~J. and {Sandford}, S.~A. and {Allamandola}, L.~J. and {Tielens}, A.~G.~G.~M. and {Sellgren}, K.},
        title = "{Near-Infrared Absorption Spectroscopy of Interstellar Hydrocarbon Grains}",
      journal = {\apj},
         year = 1994,
        month = dec,
       volume = {437},
        pages = {683},
          doi = {10.1086/175031},
       adsurl = {https://ui.adsabs.harvard.edu/abs/1994ApJ...437..683P}
}

\end{document}